\documentclass[sensors,article,submit,pdftex,moreauthors]{Definitions/mdpi} 

\firstpage{309} 
\pubvolume{26}
\issuenum{1}
\articlenumber{0}
\pubyear{2026}
\copyrightyear{2026}
\datereceived{10 November 2025} 
\daterevised{ 2 December 2025} 
\dateaccepted{16 December 2025} 
\datepublished{3 January 2026} 
\hreflink{https://doi.org/10.3390/s26010309} 
\doinum{10.3390/s26010309}

\usepackage{mathtools}
\usepackage{siunitx}
\usepackage{algorithm}
\usepackage{algpseudocodex}
\usepackage{glossaries}
\graphicspath{{figures/}{./}}

\DeclarePairedDelimiter\D{\lVert}{\rVert}
\DeclareMathOperator{\Nhx}{N}
\DeclareMathOperator{\Pos}{L}
\DeclareMathOperator*{\argmax}{arg\,max}

\newacronym{LEO}{LEO}{Low Earth Orbit}
\newacronym{ICN}{ICN}{Information Centric Networking}
\newacronym{NDN}{NDN}{Named-Data Networking}
\newacronym{ISL}{ISL}{Inter-Satellite Links}
\newacronym{PIT}{PIT}{Pending Interests Table}
\newacronym{CS}{CS}{Content Store}
\newacronym{FIB}{FIB}{Forwarding Information Base}
\newacronym{cgw}{C-Gw}{Consumer Gateway}
\newacronym{pgw}{P-Gw}{Producer Gateway}
\newacronym{OAS}{OAS}{Old Access Satellite}
\newacronym{CAS}{CAS}{Current Access Satellite}

\Title{A Scalable Solution for Node Mobility Problems in NDN-based Massive LEO Constellations}

\TitleCitation{A Scalable Solution for Node Mobility Problems in NDN-based Massive LEO Constellations}

\Author{Miguel Rodríguez Pérez $^{1}$\orcidA{}, Sergio Herrería Alonso $^{1}$*\orcidB{}
José Carlos López Ardao $^{1}$ and Andrés Suárez González $^{1}$}

\AuthorNames{Miguel Rodríguez Pérez, Sergio Herrería Alonso,
José Carlos López Ardao and Andrés Suárez González}

\isAPAStyle{%
       \AuthorCitation{Rodríguez Pérez, M., Herrería Alonso, S., López Ardao,
       J.C., \& Suárez González, A.}
         }{%
        \isChicagoStyle{%
        \AuthorCitation{Rodríguez Pérez, Miguel, Herrería Alonso, Sergio, López Ardao,
       José Carlos and Suárez González, Andrés}
        }{
        \AuthorCitation{Rodríguez Pérez, M.; Herrería Alonso; S., López Ardao,
       J.C.; Suárez González, A.}
        }
}

\address{%
$^{1}$ \quad atlanTTic research center, Universidade de Vigo}

\corres{Correspondence: sha@det.uvigo.gal}

\abstract{In recent years, there has been increasing investment in the deployment of
    massive commercial \gls{LEO} constellations to provide global
    Internet connectivity. These constellations, now equipped with
    inter-satellite links, can serve as low-latency Internet backbones,
    requiring \gls{LEO} satellites to act not only as access nodes for ground
    stations, but also as in-orbit core routers. Due to their high velocity and
    the resulting frequent handovers of ground gateways, \gls{LEO} networks highly
    stress mobility procedures at both the sender and receiver endpoints.
    On the other hand, a growing trend in networking is the use of technologies based on the
    \gls{ICN} paradigm for servicing IoT
    networks and sensor networks in general, as its addressing, storage, and
    security mechanisms are usually a good match for IoT needs. 
    Furthermore, \gls{ICN} networks posses additional characteristics that are beneficial for the massive \gls{LEO} scenario.  
    For instance, the mobility of the
    receiver is helped by the inherent data forwarding procedures in their
    architectures. However, the mobility of the senders remains an open problem.
    This paper proposes a comprehensive solution to the mobility problem for massive \gls{LEO}
     constellations using the \gls{NDN}
     architecture, as it is probably the most mature \gls{ICN} proposal. Our solution
     includes a scalable method to relate contents to ground gateways and a way
     to address traffic to the gateway that does not require cooperation from
     the network routing algorithm. Moreover, our solution works without
     requiring modifications to the actual \gls{NDN} protocol itself, so it is easy to
     test and deploy.
     Our results indicate that, for long enough handover lengths, traffic losses are negligible even for ground stations 
    with just one satellite in sight.}

\keyword{\gls{NDN}; LEO;  Mobility}

\begin{document}
\glsresetall
\section{Introduction}%
\label{sec:introduction}

During the last few years, the number of massive \gls{LEO} constellations providing
terrestrial communication services has grown dramatically. These orbiting
networks made of thousands of satellites can provide Internet connectivity to
almost any place on Earth with sky visibility. Moreover, when coupled with
\gls{ISL}, they can complement terrestrial networks, becoming
a low-latency Internet backbone~\cite{bhattacherjee_network_2019}. In these new
interconnected constellations, satellites play two simultaneous roles. On the
one hand, they are the access routers to the orbiting backbone while, on the
other hand, they are also the core routers of the
constellation transporting traffic among different geographic areas~\cite{kim_satellite_2023}.

These orbiting routers travel at very high speeds relative to their grounded
counterparts. The low altitudes of LEO orbits, just a few hundred kilometers,
yield orbital periods in the order of a hundred minutes. To accommodate these,
and to maintain connectivity, ground routers need frequent handovers between
passing-by satellites every few minutes~\cite{ye_earth_2021,wang_seamless_2023}.
Certainly, achieving end-to-end connectivity between two ground peers when part
of the routing path goes through the satellite backbone stresses common Internet
routing and mobility
protocols~\cite{yu_ip_2024,li_user-centric_2020,ma_mobility-aware_2024,dong_novel_2023}.

In traditional TCP/IP networks, that provide virtual channels for networked applications, the network and the mobility procedures have to
keep this channel stable in the face of the almost continuous topology
changes~\cite{dai_ndm_2020,hu_lisp-leo_2022}. In contrast, \gls{ICN} networks do not
rely on a virtual channel for communication. Instead, the network is able to
address \emph{data} directly and bring it to the requesting clients.

The sophisticated forwarding mechanisms of \gls{ICN} networks, along with the
motivation to explore alternative networking paradigms, have driven significant
research into their applicability as the network layer for emerging LEO
constellations. Among \gls{ICN} proposals, 
\gls{NDN}~\cite{jacobson_networking_2012} stands out for its maturity. In
particular, \gls{NDN} usage in LEO satellite networks has been a hot topic of research
lately~\cite{diao_low_2023, liu_ndn-based_2023,xie_research_2025,diao_lossless_2025}.

The modus operandi of \gls{ICN} networks, and of the \gls{NDN}
proposal in particular, inherently solves the problem of the mobility of the
receiver (or \emph{consumer} as called by \gls{NDN}). The sender (called \emph{producer})
is unaware of the consumer(s) identity and location, and the data simply travels
backwards through the same routers used by the data request. This works as long
as the path back from the producer to the receiver remains unchanged from the initial
request of the data to its final delivery. The case of producer mobility is more
involved, as it requires a method for the network to locate its current
location. \gls{NDN} networks have a standardized mechanism for this problem, the Kite
protocol~\cite{zhang_kite_2018}, but it is not appropriate for the massive LEO scenario.

Unfortunately, mobility in massive LEO constellations, with their frequent handovers
involving a great number of mobile nodes (every ground node is mobile from the
point of view of the satellite network) is not completely solved by the
aforementioned approaches. 
Producer mobility in satellite networks presents unique
challenges. Common mobility solutions, like Kite, depend on \emph{immobile anchors} (akin to
home agents in IP mobility) to maintain routes between consumers and mobile
producers. However, in massive LEO constellations, placing these anchors on the
ground is impractical, as ground nodes themselves experience frequent handovers
and cannot provide the required stability. Alternatively, assigning satellites
as immobile anchors introduces new issues: it is unclear which satellite(s)
should serve as anchors for each ground node, and the dynamic nature of
satellite orbits means that the network topology between anchors and ground
nodes changes constantly. As a result, routing through such anchors often leads
to suboptimal, longer paths and degraded performance, undermining the intended
benefits of the approach.
In the case of \emph{consumer mobility}, the return
path of data packets to the \emph{consumer} is disrupted each time the consumer
changes its access satellite, preventing the data from reaching the receiver.

In this article we propose a comprehensive solution for the mobility problem of
\gls{NDN}-based LEO satellite networks.
Our solution is able to keep connectivity between any pair of ground stations during handovers in a scalable manner and without requiring changes to the \gls{NDN} specification~\cite{named_data_networking_project_ndn_2023}.
We will obviate the routing of the requests from the consumer to the first ground station and from the second ground station to the producer, as those are already solved by the usual routing procedures.
We also provide a scalable
solution for locating the most appropriate ground node for a given producer.

Our solution solves the producer mobility problem in a scalable manner and
without relaying on \emph{home agents} or \emph{immobile anchors} like the
existing solutions in the literature. In fact, producers do not need to carry
out any extra management procedure to register their new network location after
a successful handover. This is due to the fact that the solution does not try to
be generic, but it is adapted to the specific scenario of massive LEO
constellations. Additionally, all the complexity related to the producer
movement is hidden from the satellite nodes, as they are assumed to be more
resource constrained than the ground nodes. In the same spirit, the only ground
nodes involved in the mobility procedures are the ground stations. While this is
a departure from the end-to-end principle, it simplifies the solution and eases
deployment. Finally, the proposed solution can be implemented with the facilities
present in the current \gls{NDN} version. As far as we know, our solution to the
producer mobility problem is completely novel. For the consumer mobility part of
the solution we were able to build up on some of the ideas already introduced
in~\cite{liang_ndn_2021}.

The rest of this paper is organized as follows.
Section~\ref{sec:NDN-fundamentals} provides the background and a short
introduction to \gls{NDN} fundamentals. Section~\ref{sec:problem-description}
describes the scenario and its associated mobility challenges. Then, in
Sect.~\ref{sec:producer_mobility} we show how to solve the producer mobility
problem. The consumer mobility problem is dealt with in
Sect.~\ref{sec:consumer_mobility}. Section~\ref{sec:results} shows the
experimental results. Finally, Sect.~\ref{sec:discussion} provides a discussion
on the proposed techniques and results, to lay the conclusions in
Sect.~\ref{sec:conclusions}.

\section{Basic NDN Concepts}%
\label{sec:NDN-fundamentals}

In traditional IP networks, when an IP node sends data to a peer, it
simply adds the appropriate network header with the corresponding destination IP
address to the data, and the network delivers it to the desired peer(s).
\gls{NDN} works differently. In \gls{NDN} networks, communication follows a pull-based
strategy in which \emph{consumer} applications request data from the network
that is ultimately provided by a \emph{producer} application, usually residing
on another host. This results in two different packets. The first one, demanding
a \emph{named} piece of content, is the \emph{Interest} packet that arrives to
an \gls{NDN} node through one of its interfaces (\emph{faces} in \gls{NDN} parlance). The
second one, containing the response, is the \emph{Data} packet.

\begin{figure}
    \centering
    \includegraphics[width=\columnwidth]{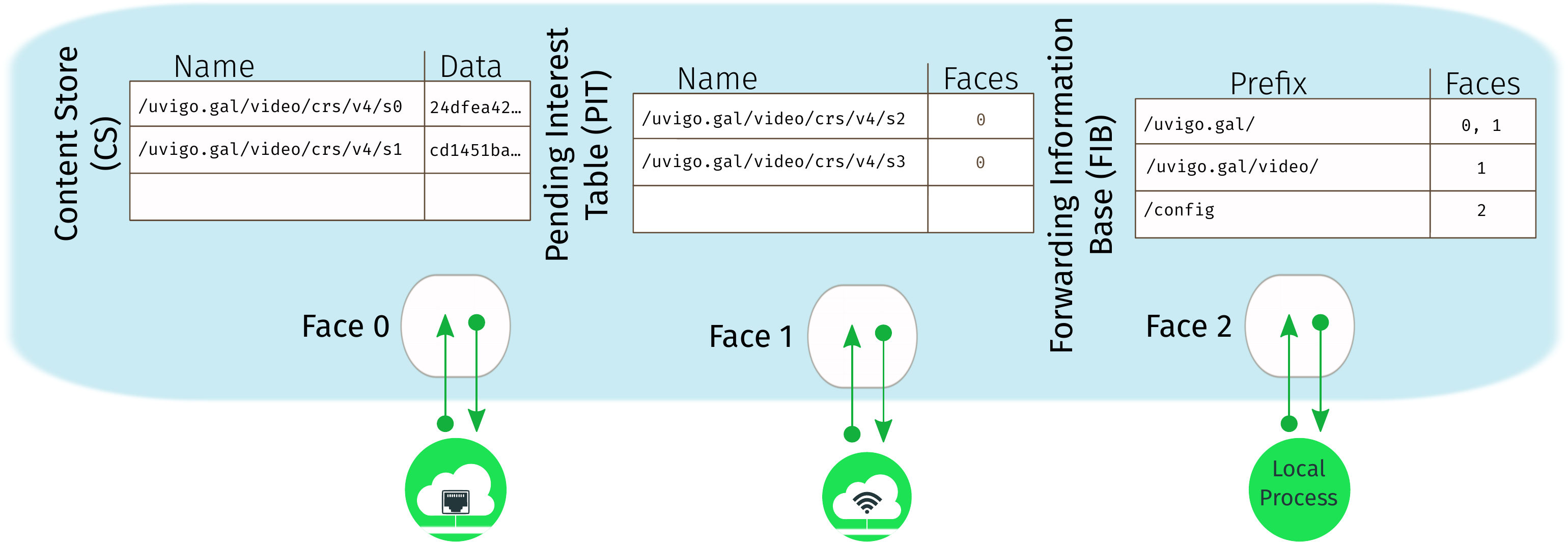}
    \caption{Main forwarding elements in an \gls{NDN} network layer.}%
    \label{fig:NDN-elements}
\end{figure}

The Interest packet contains, at least, the \emph{name} of the requested
content. \gls{NDN} nodes can implement a cache of popular contents inside their
\emph{content store (CS)} (see Fig.~\ref{fig:NDN-elements}.). Then, any \gls{NDN} node that has a copy of the requested
data (as identified by its name) can return it directly into a Data packet. To
facilitate forwarding, names in \gls{NDN} follow a hierarchical structure, so that
routers can direct the Interest packet towards a producer based on a prefix of
the name. When an \gls{NDN} router decides to forward an Interest packet, it annotates
its name and incoming interface in the \gls{PIT}.
This serves two purposes: firstly, it will be used to forward the Data packet back to
the requesting consumer; secondly, it aggregates identical requests from
different incoming interfaces, enabling seamless multicast data
transmission. To determine the outgoing face, the node uses the \gls{FIB}, which is a routing table mapping names to outgoing
interfaces. However, before indexing the table using the requested
named content, \gls{NDN} nodes use information that can be provided in the Interest packet itself.
Interest packets may carry a \emph{Forwarding Hint}, which is a list of
names associated with some topology information that are easier to route than
the actual name of the content. 

The response with the requested content travels in a \emph{Data} packet. When a
node receives an Interest and has (or can generate) the named data, it puts it
into a Data packet and forwards it to the interfaces listed in the \gls{PIT} for that
name. Intermediate nodes that receive the Data packet can store its
contents in their \gls{CS} to provide a cache for later requests
for the same name. Eventually, the Data packet reaches all demanding
consumers using the information stored in the \gls{PIT} of the intermediate
routers.

\section{Problem Description}%
\label{sec:problem-description}

Massive LEO constellations are composed of thousands of satellites organized in
shells of multiple orbital planes, each one with an equal number of equi-spaced
vessels. In recent deployments, each satellite can establish four stable
\gls{ISL}: two with the preceding and succeeding satellites
in their common orbital plane (V-ISLs) and two with the closest satellites in
the two adjacent neighboring planes in the same orbital direction (H-ISLs). Note
that these constellations mostly employ circular orbits in a Walker-delta or
Walter-start topology~\cite{noauthor_network_nodate} at a given altitude. 
As the velocity of each satellite is determined by its altitude, so it is its visibility
time from a given latitude. Thus, the minimum handover frequency decreases
with the altitude. Similarly, the number of circular orbits and the quantity
of satellites in each orbit affect the density of the constellation and, correspondingly,
the number of satellites that can simultaneously serve a given ground station.
In these designs, the
satellites in adjacent orbital planes are arranged to maintain consistent
neighboring relationships. This means that a satellite in one plane will have a
corresponding satellite in the adjacent plane that it is \emph{paired}
with~\cite{ulybyshev_near-polar_1999}. Even in more sophisticated
constellations, like Starlink, that do not follow a strict Walker-delta design,
satellites can form four stable \gls{ISL} with their closest
neighbors~\cite{jing-shi_analysis_2024}\footnote{Certainly, H-ISL links may not be
available in the polar regions if the constellation has a very high
inclination~\cite{werner_dynamic_1997}. However, we will disregard this
eventuality to keep the scenario simple, as it does not affect our mobility
solution.}.

\begin{figure}
    \centering
    \includegraphics[width=\columnwidth]{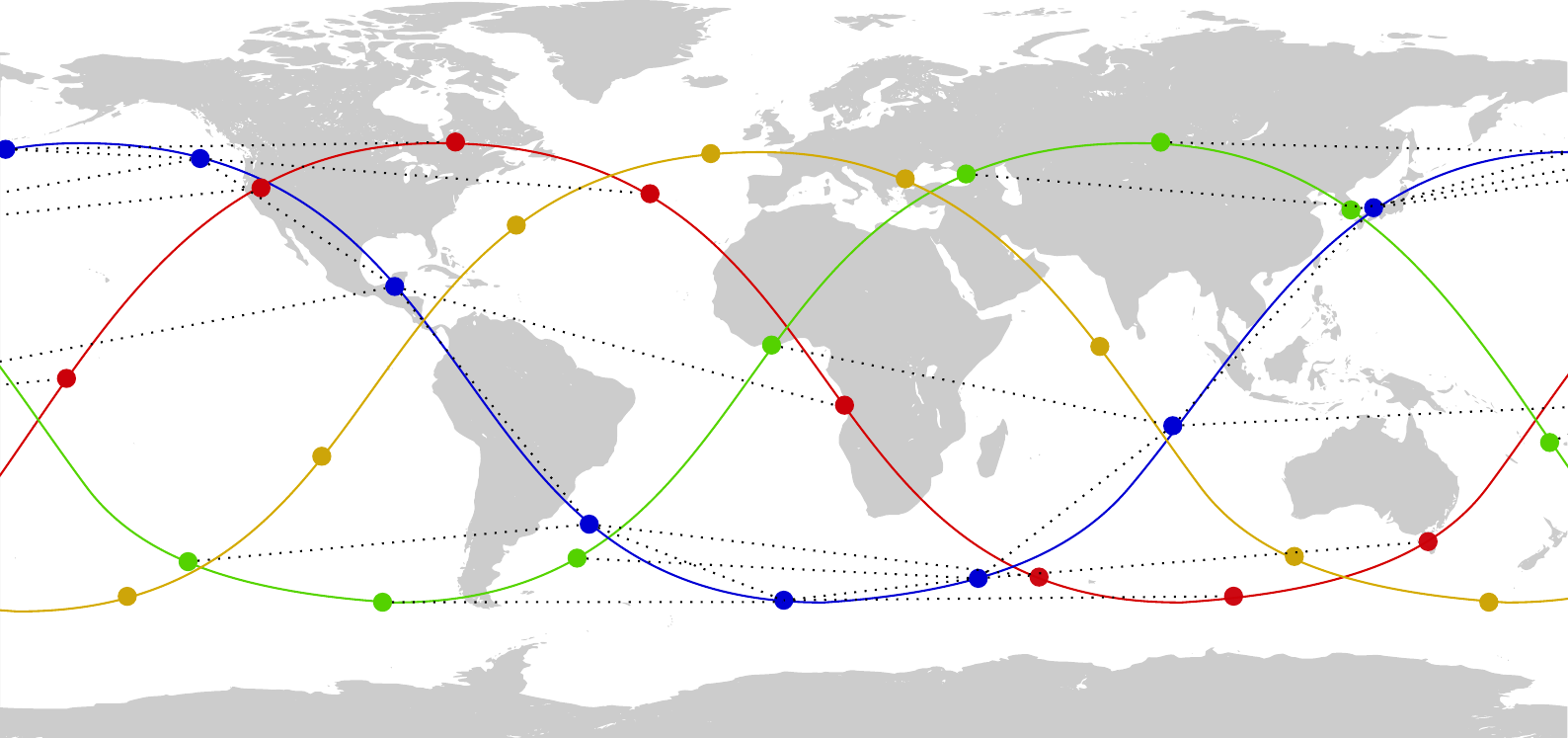}
    \caption{A small constellation with just four orbital planes and eight
        satellites per plane. The dotted lines show \gls{ISL} links of
        \emph{blue} satellites.}%
    \label{fig:constellation-isl-example}
\end{figure}

Figure~\ref{fig:constellation-isl-example} shows an example of such a
constellation.
To avoid cluttering the figure, it represents a very small constellation with
just four different orbital planes, each one with a distinct color and, with
dotted lines, the \gls{ISL} links of satellites in the \emph{blue} orbital plane. Due
to its small size, the H-ISL links shown cover long distances, but most
commercial designs for massive LEO networks are a couple of orders of magnitude
denser. Also note that the satellite in a neighboring plane may not coincide
with the closest one at a given time, since the latter may be traveling in opposite
direction. For instance, the blue satellites in the figure establish links with
the green and red satellites, but not with the yellow ones.

\begin{figure}
    \centering
    \includegraphics[width=\columnwidth]{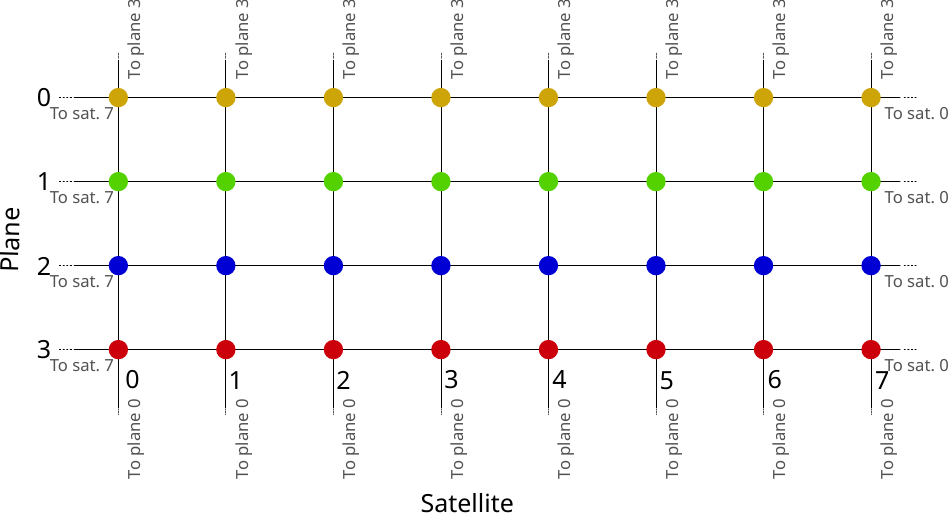}
    \caption{Logical topology of the constellation represented in Fig.~\ref{fig:constellation-isl-example}
        showing the stable links between satellites.}%
    \label{fig:constellation-isl-logical-example}
\end{figure}

The topology formed by the \gls{ISL} links does not change as the satellites orbit their
planes, because the relative locations of the satellites in their planes and the
positions of the planes remain quite stable. Thus, if we ignore the physical
location of the satellites while they travel across the globe, and focus just on
their interconnections, the actual network topology corresponds to a grid, like
the one represented in Fig.~\ref{fig:constellation-isl-logical-example}.

When all the links are operational, forwarding traffic between any two
satellites becomes straightforward, provided their grid-locations are known:
simply forwarding packets to a neighboring satellite closer to the destination
guarantees the minimum number of transmissions. However, if latency is a
concern, or when not every link is available, there are more elaborate proposals that
consider the propagation delay~\cite{tao_joint_2023} or, in \gls{NDN} networks, the
location of the caching
nodes~\cite{wu_energy-efficient_2024,rodriguez-perez_cache_2023}. In any case,
the inherent topological structure of the satellite backbone stays
relatively stable. 

The main challenge arises from the interaction
between ground nodes and their orbiting counterparts. There are frequent
handovers between ground stations and satellites, as the latter remain at sight
of a given ground location for just a few minutes at a time. Moreover, it is
unfeasible to determine which satellite(s) is(are) being tracked by a ground
station without additional information, as in a massive LEO constellation the ground station may have tens of candidate satellites to choose from.

\begin{figure}
    \centering
    \includegraphics[width=\columnwidth]{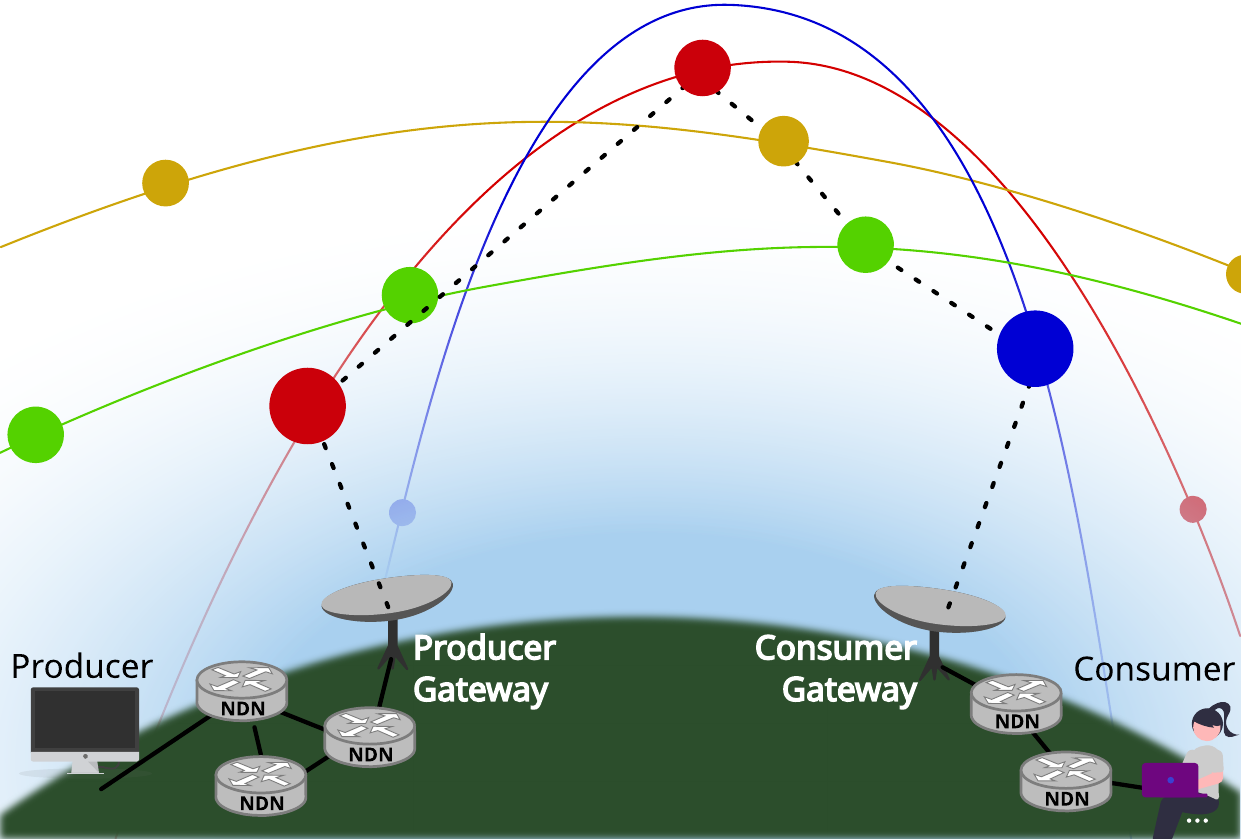}
    \caption{General overview of the scenario.}%
    \label{fig:general-scenario}
\end{figure}

We illustrate the whole transmission process with the help of the scenario depicted in
Fig.~\ref{fig:general-scenario}.
It shows  a communication path between a consumer and a producer traversing the
satellite backbone (we assume that all consumer and producers are Earth-bound). Before reaching the producer, the Interest packet sent by
the consumer must first arrive to a ground station acting as a gateway node between the ground and the
satellite network section, the \gls{cgw}. Then it traverses across the
satellite network until it reaches an appropriate ground station closer to the producer, i.e., an exit gateway that we are calling the \gls{pgw}.
Afterward, it proceeds normally towards the producer. Note that, due to the global coverage of the satellite network, there should not be more than one satellite segment in the path between the consumer and the producer. The response Data packet
follows the reverse path. We are not concerned with how the gateways are reached
by the consumer and the producer, as there is no added complexity due to the
satellite section. Instead, we will focus on how the \gls{cgw} and the \gls{pgw} can
communicate and how to identify the \gls{pgw} for a given Interest.

We assume that at any given time, both gateways can communicate with one or more
satellites, but due to the high speed of the satellites, their
respective access satellites change every few minutes.

We will first tackle how the \gls{cgw} communicates with the \gls{pgw} (producer mobility).

\section{Producer Mobility}%
\label{sec:producer_mobility}

The job of the \gls{cgw} is to forward the Interest packets it receives from
other devices to the appropriate \gls{pgw} so that they eventually reach the corresponding
producers. This entails several steps:
\begin{enumerate}
    \item The \gls{cgw} determines the identity of the \gls{pgw} for each Interest;
    \item then, it addresses the Interests to the satellite currently being used
          by the \gls{pgw};
    \item finally, the satellite network forwards the Interests to the
          appropriate satellite efficiently.
\end{enumerate}
Notice that the satellites themselves only have to deal with forwarding between
satellites, but are freed from tasks like locating the appropriate exit point.
This should remove computational load from them and place it on the ground gateways,
which have more computing and power resources.

\subsection{Identifying the P-Gw}%
\label{sec:id-p-gw}

Identifying the appropriate \gls{pgw} means obtaining the identity of the 
ground station closest to the actual producer. We name each \gls{pgw} as \texttt{<sat\_prefix>/<P-Gw>}, with \texttt{<sat\_prefix>} being the identifier assigned to
the satellite constellation operator, and \texttt{<P-Gw>} being a different identifier for each gateway.
Each ground station will be able to reach a certain amount of ground prefixes.
This information, in the form of a set of correspondences between a \gls{pgw} identity
and a prefix, must be made available to the rest of ground stations so that \gls{cgw}
can target their Interest packets accordingly.

This
information, that is not subjected to frequent variations, can be propagated with the help of a distributed database
application, similar to the DNS, and  managed by the satellite network operator.
Essentially, gateways will use this application to announce their own names
along with the set of prefixes that they can reach. When a consumer needs to
forward an Interest for a prefix, it can retrieve the name of the remote gateway
from the service. For example,
NDNS~\cite{afanasyev_NDNs_2017} and State-Vector Sync
(SVS-PS)~\cite{moll_brief_2021} are well-known methods for distributing
information on NDN networks, and both mechanisms can be easily adapted to this scenario.

\subsection{Addressing Interests to the P-Gw Access Satellites}%
\label{sec:interest-direct}

Once the identity of the proper \gls{pgw} has been determined, the next step is to
forward the Interest packet to a satellite being currently tracked by the \gls{pgw}. Then, when the Interest packet finally arrives at this tracking
satellite, it delivers the packet to the \gls{pgw} that forwards it according to the
normal procedure along the earth-bound segment of the route.

The first part of the problem is finding an actual satellite the \gls{pgw} is
tracking at the time of the Interest transmission. One possible solution would
be to use (or reuse) a distributed service mapping satellites to ground
stations. However, whereas the mapping between \gls{pgw} and reachable prefixes is
relatively stable, the relationship between ground stations and access
satellites is continuously in flux. Reusing such a system for this objective
would entail constant updates of the service database, causing excessive traffic
and computational load. Instead, we propose to encode the geographic location of
the \gls{pgw}, as the \texttt{<P-Gw>} suffix of its name. The geographic location of ground gateways is stable, and can be used to identify the set of satellites that they have
at sight at any given time. This is a straightforward task, assuming that the
\gls{cgw} has access to the constellation data. This information can be provided
in real time by the satellite network operator.\footnote{It is out of scope
how to communicate and update the satellite location data. It can certainly be provided
by the access satellites themselves to the ground station. It also should
not need high accuracy, just enough to have a rough estimate of the satellites
possible serving a given region.} 

The possibilities for encoding the node location as its suffix are plentiful, but we have decided to
rely on Open Location Code (OLC)~\cite{rinckes_open_2019} as it provides a
natural way to obtain arbitrary precision, allowing to address gateways in an
area or pinpoint their locations with as much accuracy as needed. In essence,
OLC can encode any region of the planet as an alphanumeric string of variable
length. The longer the string length, the smaller the area covered by the
region. For instance, OLC would encode the location of a \gls{pgw} located in our
campus (42.169938N, W8.687812) as \emph{plus code} \texttt{8CJH5896+XV}, with a
precision of about \qty{+-3}{\m}. So, a \gls{cgw} that just wants to contact any
gateway near the region, would address interests to
\texttt{<sat\_prefix>/8C/JH}, or near the campus to
\texttt{<sat\_prefix>/8C/JH/58}, etc.\footnote{The area covered for a given prefix length
decreases with the latitude, as the longitudinal lines become closer. \Gls{cgw} should
take this into account when determining the needed accuracy in locating a \gls{pgw}.} Then, a gateway called
\texttt{<sat\_prefix>/8C/JH/58/96/XV}, inside our building, would only answer if
the requested name contains its own name as a prefix.

The second part of the problem is getting the Interest packet forwarded to the set
of satellites instead of directly to the \gls{pgw}. Thankfully, \gls{NDN} already provides a mechanism for this. Normally, in \gls{NDN} networks, an Interest is
forwarded according to the name of the requested content. However, there exists
the possibility of providing additional information to \gls{NDN} routers as
\emph{Forwarding Hints} included in the Interest packet. These \emph{Forwarding
Hints} consist of a set of \emph{delegated} names to help the forwarding
procedure. In fact, the predefined routing \emph{strategies} use the first
reachable \emph{delegation} instead of the named content to forward the Interest
packet.\footnote{In \gls{NDN} networks the routing procedure, called \emph{strategy},
can be modified at any router and even vary according to the prefix.} In our
solution, the satellites use a modified routing strategy for the
\texttt{sat\_prefix} that employs all the \emph{delegations} listed in the
\emph{Forwarding Hint} to forward the Interest.

So, once the set of  satellites that it has at sight at any given time has been
determined, the \gls{cgw} adds their names as a list of \emph{delegations} in a
\emph{Forwarding Hint}. Additionally, the name of the \gls{pgw} is also included in the
\emph{Forwarding Hint}. This allows the actual serving satellite to identify the ground
station to which it must forward each Interest packet.

\subsection{The Forwarding Procedure}%
\label{sec:forwarding-procedure}

When an \gls{NDN} satellite receives an Interest that cannot be satisfied with its \gls{CS}
contents, it selects the follow-up forwarding actions according to the strategy
registered for the named prefix. The forwarding actions must ensure that the
Interest packet reaches all the satellites identified by the \gls{cgw} as possibly
being in use by the \gls{pgw}. However, it is also important that this forwarding
process minimizes the number of copies of the Interest packet to keep the
network load as low as possible.

We will assign a name to each satellite,
to facilitate the forwarding procedure. Since the satellite shell topology is
essentially a two-dimensional grid, each satellite can be easily identified by
its own coordinates in the grid. Thus, each satellite in the network will be
given a name of the form \texttt{<sat\_prefix>/<plane>/<index>}.
 The forwarding
procedure consists of selecting which subset of the delegation list provided by
the Forwarding Hint is going to be served by each \gls{ISL} link. Then, the Forwarding
Hint of each Interest copy forwarded to a neighboring satellite is updated to
carry just the proper subset of delegations.\footnote{This procedure even
applies to signed Interest transmissions. In \gls{NDN}, the Interest signature does
not include the Forwarding Hint, so forwarding nodes are free to alter it if
necessary without compromising the security of the communication.}
In the satellite constellation used to exemplify the mobility problem in this paper, its
grid-like topology leads to a straightforward selection of the valid \gls{ISL} links. Certainly, in the case of
more complex topologies or link-failures, a more elaborate routing algorithm should be used instead to select them.
\begin{figure}
    \centering
    \includegraphics[width=\columnwidth]{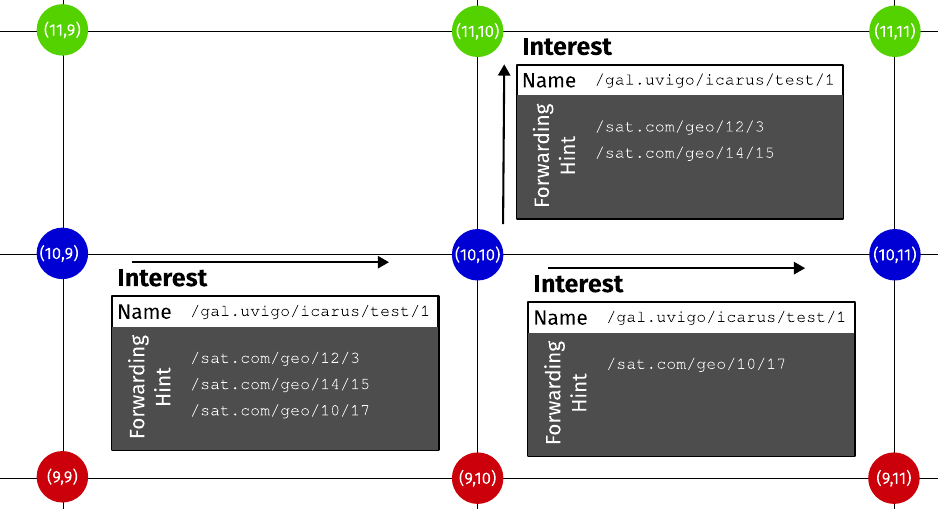}
    \caption{Interest forwarding example.}%
    \label{fig:fordward-proc}
\end{figure}
We will use the example shown in Fig.~\ref{fig:fordward-proc} to illustrate the
procedure. The Interest arriving at node (10,10) carries a delegation list
consisting of three different names, corresponding to nodes (12,3), (14,15) and
(10,17). The first step is deciding which neighbors are closer to each
destination than the current node. In this grid-like topology, this corresponds
to the 1-norm in modular arithmetic. Take delegation (12,3). Its distance from
node (10,10) is $(12-10)+(10-3)=9$, whereas from neighbor (11,10) the distance
is $(12-11)+(10-3)=8$, so (11,10) is closer to the destination, hence it is a
valid next-hop. Following the same reasoning, delegation (10,17) shall use
(10,11) as the next hop. So we need two independent transmissions for these two
delegations. Finally, delegation (14,15) has two equally valid next hops:
(11,10) and (10,11). To minimize the number of transmissions, the delegations
that can be included in more than one transmission are placed in the Interest
packet with the highest number of delegations. In this example, both carry just
one, so the selection is random. In the figure, it was included in the one
towards (11,10).

\begin{algorithm}
    \begin{algorithmic}[1]
        \State{\(\mathcal D_h \gets \emptyset,\  \forall h \in \{\mathrm F, \mathrm A,
            \mathrm P, \mathrm S\} \)}

        \LComment{Choose forwarding directions for each delegation.}

        \ForAll{\(d_i \in \mathcal D\)}
        \ForAll{\( h \in \{\mathrm F, \mathrm A, \mathrm P, \mathrm S\} \)}
        \If{\(\D{\Pos(d_i)-\Pos(s) } > \D{\Pos(d_i)-\Pos(\Nhx_h(s)) }\)}
        \State{\(\mathcal D_h \gets \mathcal D_h \cup \{d_i\} \)}
        \EndIf{}
        \EndFor{}
        \EndFor{}

        \LComment{Remove redundant transmissions.}

        \ForAll{\( h \in \{\mathrm F, \mathrm A, \mathrm P, \mathrm S\} \)}
        \If{\( \bigcup_{i \in \{\mathrm F, \mathrm A, \mathrm P, \mathrm
                S\}\setminus h} D_i \cap \mathcal D_h = \mathcal D_h \)}
        \State{\(\mathcal D_h \gets \emptyset \)}\Comment{\(D_h\) was a
            redundant set.}
        \EndIf{}
        \EndFor{}
        \ForAll{\(d_i \in \mathcal D\)}
        \State{\(\mathcal D^* \gets \{\mathcal D_h \,|\, d_i \in D_h, \forall h \in
            \{\mathrm F, \mathrm A, \mathrm P, \mathrm S\} \} \)}
        \State{\( \mathcal D^{\max} = \argmax_{\mathcal D \in \mathcal
                D^*}(\lvert \mathcal D \rvert)\)}
        \ForAll{\( \mathcal D \in \mathcal D^* \setminus \mathcal D^{\max}\)}
        \State{\(\mathcal D \gets \mathcal D \setminus d_i \)}
        \EndFor{}
        \EndFor{}
    \end{algorithmic}
    \caption{Procedure for Interest packet dissemination.}%
    \label{alg:fordward-proc}
\end{algorithm}
The exact procedure is summarized in Algorithm~\ref{alg:fordward-proc}. Its
input is the set of all the delegations present in the Forwarding Hint
(\(\mathcal D\)). The first line initializes four subsets (\(\mathcal D_\mathrm
F,\mathcal D_\mathrm A,\mathcal D_\mathrm P\text{ and }\mathcal D_\mathrm S\))
for holding the delegations to be served by each \gls{ISL} link (in the Fore, Aft,
Port and Starboard directions). Lines 3--6 implement a greedy algorithm for
selecting the appropriate forwarding set. In line 5, \(s\) represents the
current satellite; \(\Pos(\cdot)\) is the location of the given satellite;
\(\Nhx_h(s)\) is the node across link \(h\) from \(s\); and, finally,
\(\D{\cdot} \) is the 1-norm, used to calculate the distance between two nodes
in the grid-like network topology. So, at line 5, the procedure just adds
delegations to a set if the neighbor across the link is closer to the given
delegation than the current node. The second part of the procedure just removes
redundant transmissions. Lines 8--10 empty delegation sets if all their
delegations can also be carried out by the rest of the delegation sets. Lastly,
if there remain delegations served by more than one delegation set, they are
removed from all but the largest set (lines 11--15).

Eventually, when the Forwarding Hint includes the current satellite, and so as it
is a candidate to be serving the \gls{pgw}, it will forward the Interest packet
through the corresponding satellite-to-ground interface to reach the \gls{pgw}.

It is possible to adapt this procedure to route around a few possible link
failures or even to consider complete seams in the satellite topology. In the
general case, reaching the targeted satellites can be achieved with a regular
routing algorithm able to provide the appropriate next hop along the path.

\subsection{Performance Optimizations}
\label{sec:producer-performance-improvement}

When the \gls{cgw} wants to send more than a few packets to the producer, 
to reduce
unnecessary transmissions, the  \gls{cgw} should only include one actual access
satellite of the \gls{pgw} in the delegation list of the forwarded Interest packets and not every potential
satellite at its sight. This can be accomplished by previously asking the \gls{pgw} for
the name of a current access satellite. As the name of the \gls{pgw} is also known (\texttt{<sat\_prefix>/<P-Gw>}),
this can be retrieved by sending an Interest packet directly addressed to it with an
appropriate name, e.g., \texttt{<sat\_prefix>/<P-Gw>/access}. The \gls{pgw} shall respond with a
\emph{Link} object containing the name of one of its current access satellites. A Link
object is just a specialized Data packet containing a collection of names and
preferences to be used directly as a Forwarding Hint (the delegation set).
Additionally, a freshness period for this Link object is set so that it expires
before the \gls{pgw} considers a new satellite to be employed.

Since this Link object is a Data packet, it is stored in the \gls{CS} of the \gls{cgw} (and
in that of any other \gls{NDN} node in the path from \gls{pgw} to \gls{cgw} that decides to do
so). So, if during its freshness period, the \gls{cgw} sends additional Interest
packet requesting this Link object, they will be satisfied from the local \gls{CS} without
requiring a new network transmission. When the freshness period expires, any new Interest
packet for the Link object will eventually reach the \gls{pgw} to get updated
information. This provides an automatic mechanism for updating the Link object.

However, note that if the \gls{cgw} waits until the Link expires to send a new Interest to
query the access satellite, it would need to address this new Interest to every
possible satellite at sight by the \gls{pgw} again. To avoid this, we propose the following
strategy that should be valid even in the case of \gls{pgw} with single-beam
antennas limited to tracking only a single satellite at a time:
\begin{itemize}
    \item The \gls{pgw} generates the Link object with a freshness period smaller
          than the expected remaining time with the selected access satellite at
          sight. When the freshness period expires, the \gls{pgw} starts a handover process with a
          predefined duration $H$.
    \item When the Link expires in its \gls{CS}, the \gls{cgw} sends a new Interest
          requesting a new Link, using as forwarding hint the just expired
          delegation list (i.e., one with a single entry corresponding to the
          still current access satellite). This should arrive at the \gls{pgw}
          before the handover period ($H$) finishes.
    \item When the \gls{pgw} receives the Interest packet during the handover period, it generates a Link with
          two names, those of the current and next access satellites. The
          freshness period should be small, but longer than the time remaining
          until the handover period ends. This time can be equal, for example, to the handover
          length $H$.\footnote{In the case of multi-beam antennas, the link can
          contain just the name of any other satellite being tracked and the
          remaining visibility time as the freshness value.}
    \item When the handover process is complete (after $H$ seconds), the \gls{pgw}
          switches to the new satellite and assembles the corresponding new Link
          object. Shortly thereafter, the temporary Link object in the \gls{cgw}
          expires and the \gls{cgw} requests the new Link object by sending an
          Interest addressed to the names contained in the just expired one.
          This will arrive to both the old access satellite, that will not be
          able to forward it to the \gls{pgw}, and to the new one. The new access
          satellite will succeed in forwarding it to the \gls{pgw} that will answer
          with the proper Link object, valid for the next few minutes.
\end{itemize}

It must be noted that the selection of $H$ does not have to be overly precise,
only big enough to give enough time to complete the handover, but not too big
to send too many unnecessary packets simultaneously to both targets. It being a
time interval, it does not need of any synchronization between the clocks of the
participants. And, given its relatively short duration and lack of needed
precision, it should be immune to any existing clock drift. Additionally, the optimum
$H$ value, i.e, the lowest one that avoid losses, is dependent on the delay between
the \gls{cgw} and the \gls{pgw}, so, ultimately also on their locations and the
satellite segment parameters. The \gls{cgw} nodes could measure packet losses during
handovers to tune the $H$ value starting with a conservative guess.

Finally, in the case the \gls{pgw} is capable of tracking more than one satellite simultaneously,
the procedure is greatly simplified. Before one of the satellites is about to stop being
in sight, the delegation list is updated with the remaining satellites in sight and the freshness
period is set according to the satellite with less remaining time in sight.

\section{Consumer Mobility}%
\label{sec:consumer_mobility}

When a \gls{cgw} changes its access satellite, all pending Interests cannot be
satisfied because the Data packets fulfilling them cannot reach the \gls{cgw} after
the change. Even though the geographical position of the \gls{cgw} remains the same, from the
point of view of the satellite network, its point of attachment has changed,
and thus it has apparently \emph{moved}.

As mentioned in the introduction, this issue has already been addressed
by~\cite{liang_ndn_2021}. The solution presented there consists of
retransmitting pending Interests to the previous access satellite. This is also
done with the help of a Forwarding Hint. We can reuse the same approach but
performing the retransmissions directly at the \gls{cgw}, instead of at the consumer
or at any other in-network node. This satisfactorily restricts the complexity of
roaming operations to the nodes directly involved, i.e., the ground gateways and
the satellites.

\begin{figure}
    \centering
    \includegraphics[width=\columnwidth]{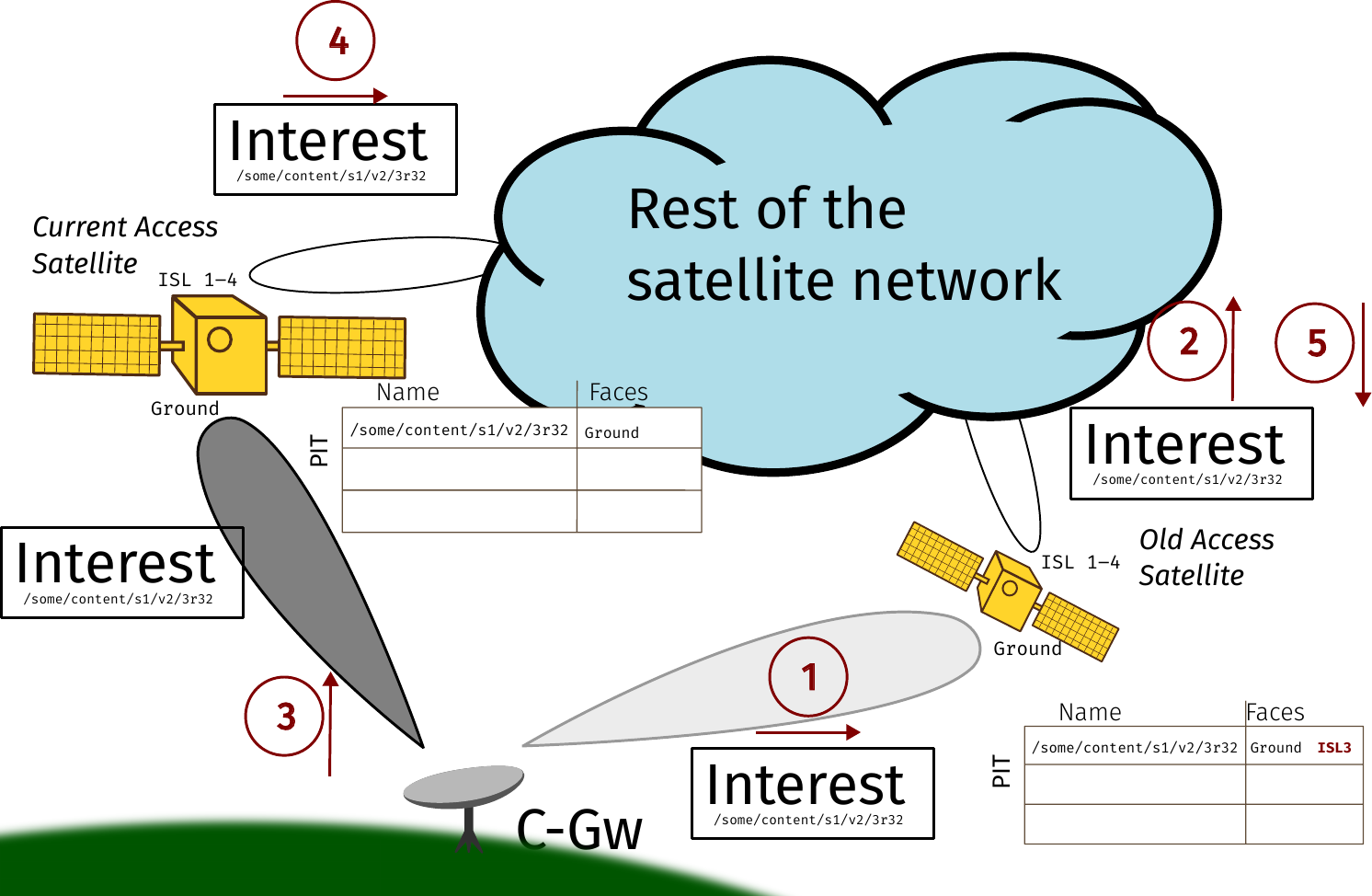}
    \caption{Consumer mobility procedure.}%
    \label{fig:cgw-roam-example}
\end{figure}

The procedure, depicted in Fig.~\ref{fig:cgw-roam-example}, is as follows.
Consider a \gls{cgw} that has some pending Interests that were transmitted via its
old access satellite (represented by steps 1 and 2 in the figure). The old
access satellite had already filled an entry in its \gls{PIT} for each pending
Interest of the \gls{cgw} through its ground interface. Then it forwards each
Interest to the rest of the satellite network through one of its \gls{ISL} links.
After a \gls{cgw} handover:
\begin{enumerate}
    \item The \gls{OAS} does not purge its \gls{PIT}\@. If the
    pending Interests are fulfilled, the corresponding Data packets are stored
    in the \gls{CS}\@.
    \item The \gls{cgw} retransmits all its pending Interests to its \gls{CAS}, but with a Forwarding Hint that employs the \gls{OAS} name
          as the only delegation. In the figure, these correspond to steps 3 to 5.
    \item The \gls{OAS} updates its \gls{PIT} adding the interface on which each Interest
    was received from the \gls{CAS} to the list in the corresponding \gls{PIT} entry. In the
    example, it is ISL3. Eventually, when it receives matching Data packets, it
    will forward them back to the \gls{CAS}, according to the information in the
    \gls{PIT}\@. Alternatively, for Data packets that have already arrived, it
    fulfills their Interests with the information from the \gls{CS}\@.
    \item The \gls{CAS} normally forwards the recovered Data packets to the \gls{cgw},
          which can then forward them to the consumers as usual according to the
          information stored in its \gls{PIT}\@.
\end{enumerate}

\section{Results}%
\label{sec:results}

To validate our approach, we have developed a simulation module for the
\gls{NDN}Sim~\cite{noauthor_NDNsim_2021} that implements the described consumer and
producer gateways~\cite{noauthor_mobility_nodate}. In particular, the code
implements 1) the required \emph{strategy} used by the satellites to forward
Interest packets according to their positions in the grid; 2) the consumer
application used by the \gls{cgw} to obtain the serving
satellite(s) of the \gls{pgw}; and 3) the producer application running on
the \gls{pgw} to respond to the Interests from the \gls{cgw}. The application at the \gls{cgw}
is also responsible for performing retransmissions of pending Interests
whenever there is a \gls{cgw} handover, as detailed in
Section~\ref{sec:consumer_mobility}. On top of this scenario, we run a stock
consumer application that asks for named data packets at a constant rate. These
Interests are satisfied by another stock producer application running on the
\gls{pgw} itself. We have not implemented, however, the service for querying the
physical location of the \gls{pgw}, since we assume that it is known beforehand.

The simulated satellite network has a design like that of the first group of the
phase of Starlink~\cite{m_wiltshire_application_2020}, with 72 orbital planes
and 22 satellites per plane, at an inclination of \ang{53} and an altitude of
\qty{550}{\km}. In each of the following experiments, the consumer and the
producer exchange data for two and a half hours (\qty{10000}{\second} to be
precise), so that there are many handovers on both the \gls{pgw} and \gls{cgw} sides. The
\gls{pgw} and the \gls{cgw} have been situated at latitude \ang{42} North, for no other
reason that it corresponds to a densely populated area on Earth, and it happens
to include the location of our lab. The distance between them is
\qty{5300}{\km}. 
With these experiments, we want to observe how the mobility
procedures affect data retrieval and the repercussions on the periods in which
the communication is interrupted. In order to try the most demanding case for
the mobility procedure, all ground stations are restricted to communicate with
just a single satellite at a time. We do not model the power constraints of the
satellites as the additional load placed on them by our mobility solution 
(the transmission of some extra Interest packets at the beginning of communications and during handovers) is negligible.

\subsection{Results at the P-Gw Site}%
\label{sec:res-p-gw}

We first analyze the results of the mobility procedures on the producer side.
Recall that every time the \gls{pgw} starts tracking a different satellite, the \gls{cgw}
has to obtain an updated delegation list that includes the identity of the new
access satellite. To ease the transition, as detailed in
Section~\ref{sec:producer-performance-improvement}, there is a handover period
of configurable length~$H$ during which the \gls{cgw} uses a delegation list
consisting of both the old and new tracked satellites.

\begin{figure}
    \centering
    \includegraphics[width=\columnwidth]{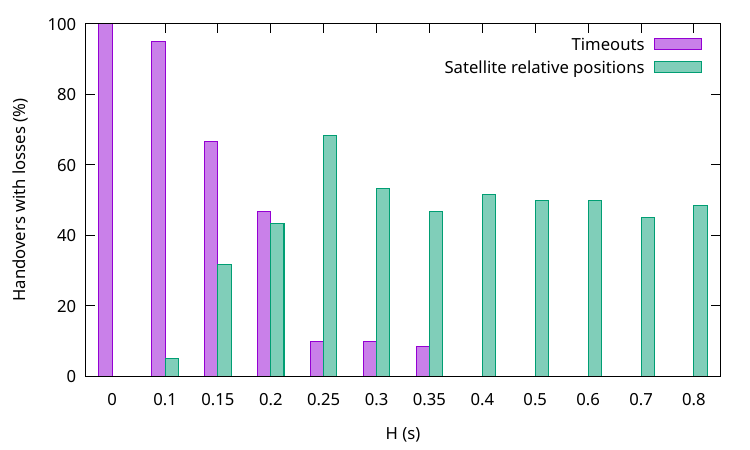}
    \caption{Percentage of \gls{pgw} handovers that suffer packet losses (and their cause) for different handover lengths.}%
    \label{fig:hist-loss-kind}
\end{figure}

Figure~\ref{fig:hist-loss-kind} shows the percentage of \gls{pgw} handovers in which
at least one Interest packet was not answered for different values of the
$H$~parameter. Losses of Interest packets can happen for two different reasons
that are distinguished in the figure. 

The first situation happens when the handover length is too small. In this case,
the \gls{cgw} will not be able to obtain the updated Link object before the \gls{pgw}
finishes the handover procedure. So, some of the Interests sent by the \gls{cgw}
will reach the old access satellite, even if the \gls{pgw} is not connected to it by
their arrival time. This will also prevent the \gls{cgw} to obtain an updated
delegation list. When the \gls{cgw} sends the Interest querying for it, addressed to
the previous \gls{pgw} serving satellite, the handover will proceed, so the Interest
will not ever reach the \gls{pgw}. This cannot be avoided, not even by sending this
Interest before. If the \gls{cgw} were to send the Interest before the handover
happened, by the time the Interest reaches the \gls{pgw}, the handover process would
not have been started yet, and so the Link object will consist of the satellite
to be abandoned. When this happens, the \gls{cgw} will keep on sending Interests to
the old access satellite until they timeout. In any case when the stranded
Interest querying for the updated delegation list timeouts (\qty{1}{\second} in
our experiments to see it easily) the \gls{cgw} will calculate the complete set of
possible access satellites, obtaining a new delegation list, and the
communication will resume, albeit in a less efficient manner until the identity
of the new access satellite can finally be determined, as per
Section~\ref{sec:producer-performance-improvement}. In
Fig.~\ref{fig:hist-loss-kind}, these traffic loss periods that end after a
timeout are responsible for most of the traffic loss periods when the handover
length is small. In fact, when the handover length is \qty{0}{s}, all handovers
suffer packet losses due to this cause. As the handover length increases, the
percentage of handovers leading to timeouts diminishes. Indeed, timeouts occur
only when the transmission delay between the \gls{cgw} and the old access satellite is less
than the handover length. On the right side of the graph, when the handover
length is high enough, the \gls{cgw} is always able to get the updated Link object
and it never timeouts.

The second kind of losses depends on the relative distances, from the point of
view of the \gls{cgw}, to the old and new access satellites of the \gls{pgw}. Note that
Interest packets get lost when the new access satellite is closer to the \gls{cgw}
than the old one. As this is random, it approximately happens in
\qty{50}{\percent} of the handovers, as shown in the figure. To see how this
occurs, consider an Interest that is being directed at both the old and the new access
satellite. When the new one is closer to the \gls{cgw}, the Interest should reach it
before it gets to the old one. However, for a few Interests this happens
\emph{before} the handover has been completed, and so they are discarded by the
yet-to-be access satellite. Then, if the actual handover is finalized just after
this, and before these Interests reach the old access satellite, they are
discarded by both. Note that, even though this situation can potentially happen
in half of the mobility events, it does affect only a few packets, since the
duration of the event is very short.

\begin{figure}
    \centering
    \includegraphics[width=\columnwidth]{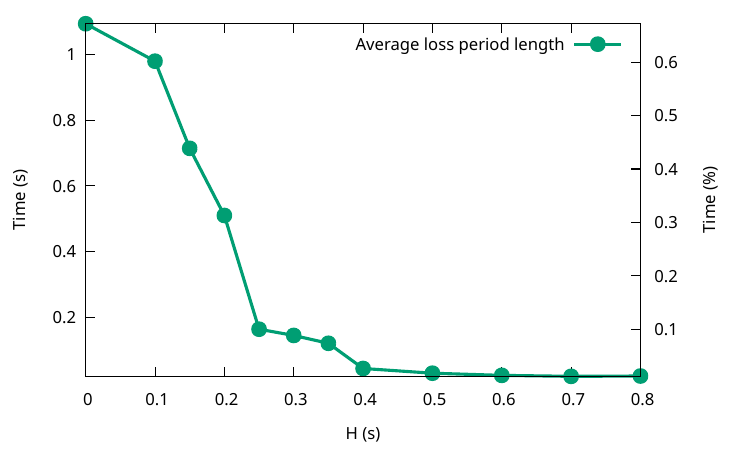}
    \caption{Average length of loss periods after \gls{pgw} satellite
        handovers.}%
    \label{fig:av-gap-length-producer}
\end{figure}
To shed more light on the impact of \gls{pgw} handovers on traffic loss, we also
measured the average length of the traffic loss period in each handover for
different values of the handover length parameter. The results are shown in
Fig.~\ref{fig:av-gap-length-producer}, both as the absolute time (left axis) and
as the percentage of the time between two handovers (right axis). For small
handover lengths, when most losses are due to timeouts, the loss length is close
to the timeout value. In fact, without a handover period (i.e., $H=0$), it is
slightly higher, because this length includes both the timeout and the time to
get a new valid delegation list. However, as the handover length increases and
fewer and fewer loss periods are due to timeouts, the average length of the loss
periods gets closer to 0. This corresponds with the loss events due to the
relative positions of the old and new access satellites and, in these cases, the
length of the loss periods is limited to the difference between their relative
positions, that is usually very small. Notice also that, for any $H$ value, the
time period when losses may happen is always less than \qty{1}{\percent} of the
time between consecutive handovers.

\subsection{Results at the C-Gw Site}%
\label{sec:res-c-gw}

The consumer mobility procedure also affects traffic delivery. Whenever the \gls{cgw}
changes its access satellite, the return path of the Data packets serving
pending Interests is broken. When the consumer mobility procedure detects the
mobility event, it recovers the missing data re-sending the Interests through
the new access satellite using a forwarding hint to route the Interests to the
old access satellite. If the Data arrived to the old access satellite before the
retransmitted Interest reaches it, it answers with a copy of the Data from its content
store. If not, the incoming interface is added to the \gls{PIT} entry to the requested
content. When the Data finally arrives to the old access satellite, a copy is sent back
to the new one and then arrives to the \gls{cgw}.

To exemplify how this mechanism performs we have simulated a scenario where a
consumer sends one thousand Interests per second to a producer in the same LEO
network topology as before. The purpose of the traffic rate is not to measure
the network capacity, which is orthogonal to the mobility issue, but to capture
the times when packets are either delayed, retransmitted or lost due to
mobility.
\begin{figure}
    \centering
    \includegraphics[width=\columnwidth]{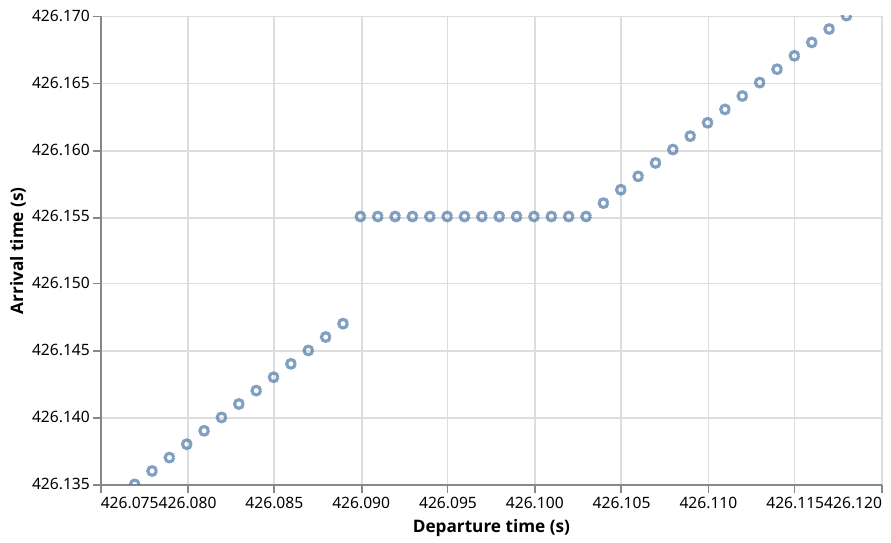}
    \caption{Departure time of Interests and arrival time of the corresponding
    Data to the \gls{cgw} during a consumer mobility event. Each circle represents
    the arrival of a Data packet.}%
    \label{fig:mobility-simulation-cgw}
\end{figure}
Figure~\ref{fig:mobility-simulation-cgw} shows the arrival time of Data packets
and the departure times of their corresponding Interest packets. At time
\qty{426.148}{\second} the access satellite is modified. Immediately there is a
gap in the arrival times, since the Data packets queued for transmission to the
\gls{cgw} at the old access satellite cannot be transmitted. The \gls{cgw} mobility
procedure requests again old pending data in the \gls{PIT} of the \gls{cgw} and, at time
\qty{426.155}{\second} the re-requested packets arrive through the new access
satellite. They arrive almost simultaneously because they had all already been
received by the old access satellite, so, in this case, the round trip time
between the \gls{cgw} and the old access satellite via the new one is just
\qty{7}{\ms}.\footnote{An actual implementation may want to pace the
retransmission of pending Interests to avoid traffic surges in the return path.}
Note that all Data packets have been recovered, as there are no horizontal gaps
between consecutive packets. Finally, at time \qty{426.156}{\second} the first
non retransmitted request arrives, and the transmission goes on normally.

\section{Discussion}%
\label{sec:discussion}

Any solution to the mobility problem of massive satellite communications must be
both scalable and secure.

There are three distinct elements that participate in our solution: the consumer
gateway, the producer gateway, and the satellites themselves. The strategy we
have chosen to map \gls{NDN} prefixes to \gls{pgw} is akin to the one used on the Internet
to map domain names to IP addresses, so scalability should not be a concern.
Regarding the challenge of mapping the \gls{pgw} to the set of possible satellites at
its sight, identifying these satellites does not involve any communication since
it is carried out by the \gls{cgw} using the ephemeris information of the
constellation. However, the size of this set grows linearly with the
constellation size. We do not believe that this is a serious issue, as only the
first Interest packet between a \gls{cgw} and a \gls{pgw} addresses the whole set of
possible satellites. Once the actual satellite has been determined, the rest of
Interest packets address only one (sometimes two) satellites, regardless of the
constellation size. Finally, once an Interest reaches the  LEO segment of the
path, routing towards the exit satellite is trivial due to its grid topology,
and therefore the complexity is independent of the network size.

Security issues must be also considered, as all traffic travels through
intermediate nodes that can inspect it. Fortunately, \gls{NDN} design already
considers these concerns: every Data packet is cryptographically signed by the
producer, ensuring integrity; additionally, their content can also be ciphered
at the producer, thus providing confidentiality. Moreover, even the suffix parts
of the name of the requested data can be ciphered so that only the producer
application can decode it.

\section{Conclusions}%
\label{sec:conclusions}

Maintaining uninterrupted communication between ground terminals across a
massive LEO constellation is challenging for conventional Internet technology
due to frequent handovers. This article explores the feasibility of utilizing
the \gls{NDN} architecture to address this issue without requiring modifications to
the user software or the \gls{NDN} protocol itself.

We presented an efficient, scalable method for identifying satellites serving
specific geographic locations in order to drive traffic back to ground nodes close to the producers.
The method includes the means to reduce the number of satellites receiving the traffic for a given ground node
to the actual set being tracked by the ground node.

Additionally, we demonstrated the adaptability of gateways to respond
to remote producer mobility while minimizing traffic losses and also successfully
adapted an existing consumer mobility protocol for our scenario to also address consumer
mobility. When taken  together, the method for identifying satellites and the 
produced and consumer mobility solutions provide a complete solution for node mobility
in \gls{NDN} massive \gls{LEO} constellations.

The results show that our producer mobility solution significantly reduces
traffic losses by forwarding Interest packets to two satellites during handover
periods, which can be as short as half a second.

Future lines of research need to address the routing of Interest packets in satellite shells with a grid-like topology in more realistic scenarios, for instance, 
considering link failures and/or incomplete networks. We believe that a low-complexity
routing solution can be provided due to the regularity of the scenario. A second future research line is the design of a dynamic method for tuning the $H$~parameter that minimizes the handover lengths while keeping the packet losses controlled.

\authorcontributions{Conceptualization: MR and SH; Formal Analysis: AS; Funding acquisition: MR and SH; Investigation: MR; Methodology: JCL; Software: MR; Supervision: SH and JCL; Validation: AS; Writing - original draft: MR; Writing - review \& editing: SH, JCL and AS.}

\funding{This work has received financial support from grant PID2020-113240RB-I00, financed by MCIN/AEI/10.13039/501100011033.}

\institutionalreview{Not applicable.}



\dataavailability{Dataset available on request from the authors.}

\conflictsofinterest{The authors declare no conflicts of interest.} 

\begin{adjustwidth}{-\extralength}{0cm}

\reftitle{References}

\bibliography{mobility}

\PublishersNote{}
\end{adjustwidth}

\end{document}